\documentclass[runningheads]{svmult}

\usepackage{makeidx}   
\usepackage{graphicx}  
\usepackage{subeqnar}  
\usepackage{multicol}  
\usepackage{physprbb}  
\makeindex             

\newcommand{\greeksym}[1]{{\usefont{U}{psy}{m}{n}#1}}
\newcommand{\umu}{\mbox{\greeksym{m}}}

\usepackage[colorlinks=false,bookmarks=false,pagebackref=false]{hyperref}

\hypersetup{
  pdftitle={Brown Dwarfs and Hot Jupiters},
  pdfauthor={Derek Homeier}}
\begin{document}
\errorcontextlines=\maxdimen
\title*{Spectral Properties of Brown Dwarfs \protect\newline and Hot Jupiters}

\toctitle{Spectral Properties of Brown Dwarfs and Hot Jupiters}
%
%
\titlerunning{Brown Dwarfs and Hot Jupiters}
%
\author{Derek Homeier\inst{1}
\and France Allard\inst{2}
\and Peter H.~Hauschildt\inst{3}
\and Travis S.~Barman\inst{4}
\and Andreas Schweitzer\inst{3}
\and Edward A.~Baron\inst{5}}
\authorrunning{Derek Homeier et al.}
%
%
\institute{Department of Physics \& Astronomy, University of Georgia, 
 Athens, GA 30602-2451, USA; derek@physast.uga.edu 
\and Centre de Recherche Astronomique de Lyon, 
 {\'E}cole Normale Sup{\'e}rieure, 69634 Lyon Cedex 07, France 
\and Hamburger Sternwarte, Universit{\"a}t Hamburg, 
 21029 Hamburg, Germany 
\and Department of Physics, Wichita State University, 
 Wichita, KS 67260-0032, USA
\and Dept.\ of Physics \&\ Astronomy, University of
Oklahoma, Norman, OK 73019, USA}

\maketitle              

\thispagestyle{myheadings}
\markright{to appear in \quad\em High Resolution Infrared Spectroscopy
  in Astronomy \hspace{4in}}
\begin{abstract}
{Brown dwarfs}\index{brown dwarfs} bridge the gap between the stellar and planetary
mass regimes. Evolving from conditions very similar to the lowest-mass
stars, the atmospheres of older brown dwarfs closely resemble 
those expected in close-in {extrasolar giant planets}\index{extrasolar giant planets}, and with
cooler BDs still being discovered, more and more approach the
properties of gas giants at wider separation. Interpreting the spectra
of BDs is therefore a crucial step towards understanding and
predicting the spectral and thermal properties of EGPs. 

Essential properties of substellar atmospheres are massive molecular
line-blanketing and the condensation of species with decreasing
{$T_{\mathrm{eff}}$}, changing the chemical equilibrium composition
and causing absorption from dust grains. 
More complex details involve the distribution of dust clouds over the
surface giving rise to temporal variability, and possible deviations
from chemical equilibrium conditions. In the case of close-in EGPs and
some BDs in binary systems, the effect of irradiation from the primary
significantly affects the spectral properties and thermal
evolution. 
\end{abstract}

 
\section{Introduction}
Observational efforts during the past decade have brought marked
progress in characterising the lower end of the Main Sequence 
down to and beyond the hydrogen-burning limit, and identifying a class
of unambiguously {substellar}\index{substellar} objects. Brown dwarfs (BDs) are
commonly defined as compact objects with a mass below the minimum for
sustaining equilibrium hydrogen 
fusion ($\sim\!0.07 M_\odot$). 
Since the detection of the first bona-fide BDs
\cite{reboloPleiade,nakaGl229B,broGl229B} and the discovery of a 
Jupiter-mass companion to 51\,Peg by Mayor \& Queloz
\cite{mayor51Peg}, the field of substellar 
astronomy has seen the direct detection of more
than 300 ultracool dwarfs and the indirect detection of 110 
planetary mass objects in orbit around other stars. 
While there is no agreement yet whether brown dwarfs should be
distinguished from extrasolar giant planets (EGPs) based on their
formation history, or by defining the minimum mass for {deuterium
burning}\index{deuterium burning} ($\sim\!13 M_\mathrm{J}$) as the
lower limit of the BD mass 
range, in terms of atmospheric properties and spectroscopic appearance
there is a smooth transition, and some overlap, between both classes
of objects, regardless of definition. 

The spectroscopic characterisation of these sources 
of 2200\,K\,$>T_\mathrm{eff}>$\,700\,K has prompted the introduction
of the two new {spectral classes L and T}. While the original
description of the L sequence was based on optical spectral properties 
\cite{mbdfLdwarf97,krlLdwarf99}, the definition of a classification
scheme for the latest L-type\index{L dwarfs} and the 
{T dwarfs}\index{T dwarfs} has 
required the use of near-IR spectra, owing to the 
fact that the lion's share of emitted flux and of spectroscopic
characteristics in these objects is found at $\lambda>0.95$\,\umu m 
\cite{geb02,adam02a}. 

While the youngest / most massive brown dwarfs jointly populate the
early L (and late M) classes together with the least massive
hydrogen-burning (very low mass-, or VLM-) stars, more evolved brown
dwarfs of late-L type  
have effective temperatures similar to those of the EGPs closest-in
to their primary star (\lq \lq {Hot Jupiters}\index{Hot Jupiters}", 
e.\,g.\ 51~Peg, HD~209458 \cite{evolPlanets}). 
The coolest T dwarfs detected to date \cite{tomgGl570,2MASS04pap}
exhibit atmospheric properties close to those anticipated in EGPs at
larger separation, and thus are crucial test cases for our
understanding of planetary atmospheres.  

The spectral appearance of substellar objects, proceeding from the
lowest-mass main-sequence stars to lower effective 
temperatures is mainly characterised by the onset of condensate
formation in the latest {M dwarfs}\index{M dwarfs}, leading to strongly
dust-dominated atmospheres of early L-types with very red
$J\!H\!K$-colours. From mid-L to early T-types this trend is reversed
with a decreasing signature of dust, strengthening of water vapour
absorption and the appearance of methane. The strongest CH$_4$ bands
at 3.3 and 2.2\,{\umu m} have  been observed 
in dwarfs as early as L5 \cite{2000ApJ...541L..75N}. The L/T 
transition is thus now defined by the first appearance of the
weaker band at 1.6\,{\umu m} \cite{geb02,adam02a}. 
While the near-IR spectra of T dwarfs are turning blue
again with decreasing $T_\mathrm{eff}$, caused by the deepening
molecular bands and 
{collision-induced absorption}\index{collision-induced absorption}
(CIA) of H$_2$ in the $K$-band, their optical-infrared colours become
extremely red due to the massively pressure-broadened 
resonance lines of Na\,I and K\,I.  

Chemical interaction, the complex spectra of molecules, the
physics of {dust formation}\index{dust formation} 
and the treatment of {line broadening}\index{line broadening} in high density
conditions make spectral modelling of brown dwarfs and giant planets a
challenging task.  
With advancements in computational techniques and availability of
better chemical and physical input data over the years, models of the
atmospheres and interiors, 
\cite{1993ApJ...406..158B,faphh95,tsujiDust96,evolDust} 
have been able to keep up with observational progress in their 
ability to describe the global properties of brown dwarfs. 
In this review, we present an introduction into 
dwarf atmosphere modelling with the \texttt{PHOENIX} code in
Sect.~\ref{phoenix}, followed by an overview of the current status of
models and recent developments in Sect.~\ref{status}. 
We discuss the special case of modelling the 
{irradiated atmospheres}\index{irradiated atmospheres} of close-in
extrasolar giant planets in Sect.~\ref{planets} and discuss the
outlook to possible direct observations of EGPs.  

\markboth{D.~Homeier et al.}{Brown Dwarfs and Hot Jupiters}
\section{The \texttt{PHOENIX} Code}\label{phoenix}

\href{http://www.hs.uni-hamburg.de/phoenix/}{\texttt{PHOENIX}} 
(see \cite{hbjcam99} and references therein) is a
general stellar model atmosphere code for treating both static and moving
atmospheres, designed to be both 
general enough to allow essentially all astrophysical objects to be
modeled with a single code, and to make as few approximations as possible. 
Successful applications of the \texttt{PHOENIX} code include models of 
Novae, all types of Supernovae \cite{blhSNIa03} and
Hot Stars \cite{deneb}.  
Compared to these cases {cool stars}\index{cool stars} and 
{brown dwarfs}\index{brown dwarfs}
are in many respects simpler instances of the classical stellar
atmosphere problem, i.\,e.\ the photospheres can be considered as
plane parallel and the assumptions of hydrostatic equilibrium, 
{local thermodynamical equilibrium}\index{local thermodynamical
  equilibrium} (LTE) and flux conservation at each level hold to a
high degree.  
Complications arise from the need to treat
{convection}\index{convection} into
optically thin layers, highly non-grey opacities that cause extreme
deviations of the radiation field from that of a blackbody, 
and the prevalence of
{molecules}\index{molecules} at low temperatures. 

\subsection{Atmospheric Structure and Chemistry}\label{structchem}
Molecules first appear at
temperatures of about 5\,000\,K, and dominate the conditions 
in M dwarfs, locking up hydrogen and most of the metals. 
This dramatically increases the complexity of the 
{equation of state}\index{equation of state} (EOS) as chemical interactions of
hundreds of species have to be considered simultaneously. 
The \texttt{PHOENIX} code currently solves for the 
{chemical equilibrium}\index{chemical equilibrium} (hereafter CE) 
partial pressures of 40
elements with usually 2 to 6 ionization stages each, 
over 600 molecular species \cite{faphh95,LimDust}, and  
over 1000 liquids and crystals from a 
study by Sharp \& Huebner \cite{Sharp+Huebner90}. 
Thermodynamic equilibrium models, however, can only describe the
conditions, under which condensates {\em may} be formed, but not 
the actual nucleation rate and growth of grains. 
Therefore, two limiting cases for the presence of dust in the
photosphere have been described by Allard et al.\ \cite{LimDust},
referred to as AMES-Dusty and AMES-Cond. 
While in both cases dust is assumed to have
formed wherever it can under CE, removing refractory elements  
from the gas phase, the former models consider 
all grains to remain at the place where they formed, while the 
latter assume that dust will completely settle to deeper layers
and does not contribute to the opacity at all. 

\subsection{Opacities}\label{opacity}
With the electronic, vibrational and rotational transitions of
molecules being orders of magnitude more complex than atomic line
spectra, the handling of up to several $10^8$ line opacities 
poses a computational problem of its own. 
The input data size is illustrated by a breakdown of lines
in our opacity database: 
\begin{itemize}
 \itemsep=1pt
\item number of atomic/ionic spectral lines: $\approx 42\times 10^{6}$ $\to 0.6\,$GB

\item number of diatomic molecular lines (other than TiO): $\approx 35\times 10^{6}$ $\to 0.5\,$GB

\item number of TiO lines (1999): $\approx 170\times 10^{6}$ $\to 2.5\,$GB

\item \begin{tabbing} number of hot water vapour lines
\= -- before 1994: \=\quad\= $\approx 0.035\times 10^{6}$ \= \\
\> -- 1994: \> $+$\> $\approx 6\times 10^{6}$   \> $\to 0.1\,$GB\\
\> -- 1997: \> $+$\> $\approx 308\times 10^{6}$ \> $\to 4.5\,$GB\\
\> -- 2001: \> $+$\> $\approx 100\times 10^{6}$ \> $\to 1.5\,$GB
  \end{tabbing}
\item total molecular lines (Oct 2001): $\approx 700\times 10^{6}$ $\to 10\,$GB
\end{itemize}



{Line blanketing}\index{line blanketing} is treated with the 
{\em direct opacity sampling}\index{direct opacity sampling} method, by dynamically  creating
sub-lists of all transitions contributing to the opacity for the
conditions of a given model, and calculating depth dependent Gauss
(weak lines) or Voigt (strong lines) profiles \cite{jcam}. For
important resonance lines, such as the strongly broadened alkali D1/D2
lines, special damping profiles \cite{Alkalis03} are used. 

\begin{figure}[htbp]
\begin{center}
\includegraphics[width=.64\textwidth]{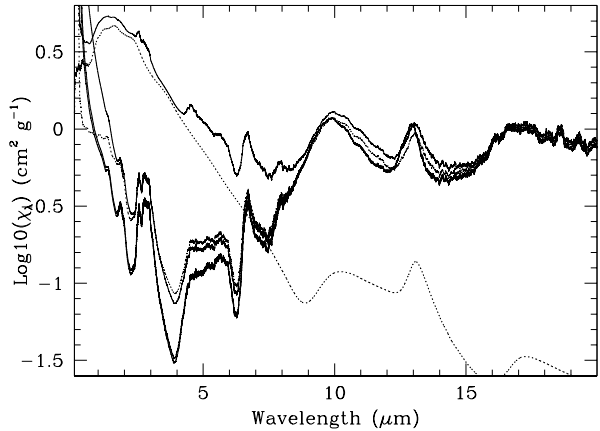}
\end{center}
\caption[]{\label{gsize}  Combined dust  extinction  profiles for 
  power-law grain size  distributions with 1, 2,  10 and 100 times
  the standard ISM values (full lines from  bottom to  top
  respectively, where  the  two  first  curves  are  nearly
  undistinguishable).  The scattering  and absorption contributions
  of the  100 ISM  profile are also  shown   (dotted  lines).   The
  conditions are typical of the upper photospheric layers
  in a 1\,800\,K AMES-Dusty model atmosphere (see Allard et
  al.~\cite{LimDust} for further details). 
  The structures seen in the profile at $\lambda >  8.5\mu$m are due to
  dust absorption (Mg$_2$SiO$_4$ at 10 and 16.5 $\mu$m and MgAl$_2$O$_4$
  at 13 $\mu$m).  Scattering contributions dominate below $0.5\mu$m,
  but remain modest in the IR for all but the largest grain sizes. 
  The absorption profile shows little sensitivity to grain sizes. }
\end{figure}

For models where {dust opacity}\index{dust opacity} is considered (AMES-Dusty), 
extinction from 31 grain species are currently included, 
with absorption and scattering cross-sections calculated using
the {Mie formalism}\index{Mie formalism} applied to spherical grains 
\cite{draAtmos}. 
Figure~\ref{gsize} shows the combined extinction from all dust species
included in our EOS, which provide an
essentially continuous opacity, though several pronounced 
features exist in the mid-IR, due to e.\,g.\ the absorption peaks of
silicates and corundum (Al$_2$O$_3$). 
An uncertain factor in these calculations is the
size of the grains, about which only little is known for the
conditions of 
dense stellar atmospheres. For lack of a better model, a power-law
distribution is frequently assumed, adopting an ISM model of 
$n(a) \propto a^{-3.5}$ for grain diameters 6.25\,nm\,$\le\! a
\!\le$\,250\,nm. However, the total extinction per dust mass remains
insensitive to particle size as long as the grains remain within the
{Rayleigh limit}\index{Rayleigh limit} $a \ll \lambda/2\pi$, and 
therefore the IR
opacity would only be affected if the grains became much larger than in
the ISM (cf.\ Sect.~\ref{settling}). 

\section{Current Status of Models}\label{status}

\begin{figure}[t]
\begin{center}
\includegraphics[width=.96\textwidth]{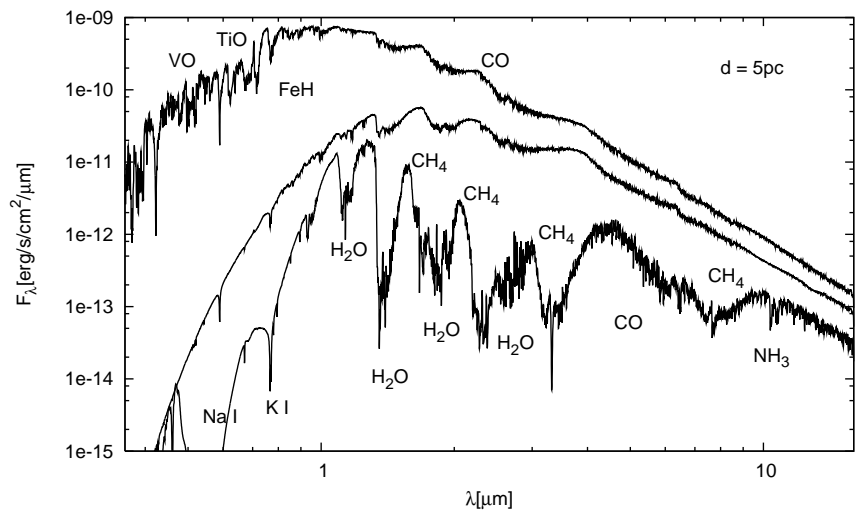}
\end{center}
\caption[]{Model spectra for the conditions typical of a VLM star
  ($T_\mathrm{eff}=3\,500$\,K, top, with very little dust being
  formed), a young BD 
  ($T_\mathrm{eff}=1\,800$\,K, middle, with dust opacity included) and
  an older field BD 
  ($T_\mathrm{eff}=1\,000$\,K, bottom, without dust opacity). The most
  important absorption features are indicated.}
\label{MLTspectra}
\end{figure}

The evolution of spectral properties from the hydrogen-burning limit
to the coolest observed brown dwarfs is displayed in
Fig.\,\ref{MLTspectra}. The transition from M- to L-types is
characterised by disappearance of the optical VO and TiO bands, while
FeH  becomes stronger in the red optical and near IR spectrum, and a
shift of the IR SED, exhibited in increasingly red ($J\!-\!K$)
colours. These changes are explained by depletion of gaseous TiO due to
the formation of titanates and to a
lesser degree the condensation of VO, and the added
opacity of silicate grains. Dust extinction blocks 
efficiently the emission in the optical and tends to smear out
the remaining molecular absorption bands, the most prominent being 
due to H$_2$O and CO. 

Dusty models such as the middle plot in Fig.\,\ref{MLTspectra}
reproduce well the observed spectra of early L dwarfs. 
As $T_\mathrm{eff}$ sinks below 1\,700\,K, silicates form at 
deeper layers, where they affect 
the emerging spectrum less, the outermost layers can cool and
molecular bands begin to grow in strength again. The cooling also
triggers a rapid change from a CE composition where most carbon is
bound in carbon monoxide to a carbon chemistry dominated by methane,
allowing the much stronger bands of this molecule to shape the IR
spectrum. We can see the end point of this transformation at 
$T_\mathrm{eff}$\,=\,1\,000\,K, with the bottom plot showing an
AMES-Cond model where the atmosphere is dust-free, and 
water and methane bands are strong. Consequently, cool evolved
brown dwarfs emit more flux at shorter wavelengths (in the near-IR 
$Z$, $J$, and $H$ broad bands), between the strong water vapour and
methane bands, than blackbodies of the same bolometric luminosity would
predict, and show very blue ($J\!-\!K$) colours compared to L 
dwarfs. This important fact was already predicted by the model
calculations of Allard et al.\ in 1995 
\cite{faphh95} and confirmed by the discovery of the first evolved
brown dwarf, Gliese 229\,B \cite{nakaGl229B,broGl229B}.

\subsection{Formation and Gravitational Settling 
  of Dust Clouds}\label{settling} 

Therefore, based upon observational evidence, the latest
M and most {L dwarfs}\index{L dwarfs} carry atmospheric dust, whereas late
{T dwarfs}\index{T dwarfs} show
little or no evidence of condensates contributing to the opacity.  
Our models presented in Sect.~\ref{structchem} reproduce these
extreme states well, but cannot adequately describe the transition
between the two. 
In fact, in a static equilibrium model there is no solution that 
allows dust to remain suspended in the photosphere at all, as the 
grains will eventually sink to lower layers, and take all condensible
elements with them until their abundance has decreased below the
saturation pressure. 
Radiation pressure is negligible at such low luminosities and
certainly cannot provide the force required to keep grains lifted up. 
Because of the relatively inefficient transport of energy by radiation
and the large {molecular opacities}\index{molecular opacities}, 
these cool atmospheres are all
convective. {Convection}\index{convection} can thus provide 
the essential replenishment of condensible material in the cloud
layers.   We therefore understand the presence of dust as a
delicate equilibrium between 1) the sedimentation of the largest grains
by gravity to deeper, hotter layers where they are melted and
sublimated, 2) the convective updraft of condensible material, and 3)
the efficiency of the condensation process and growth rate of
grains. It becomes clear  from this that the cloud layer can only
exist in close proximity to a convection zone, where mixing
is still vigourous enough to balance the sedimentation and depletion
process. 
As the radiative-convective boundary retreats with decreasing
luminosity, the cloud top also sinks deeper into the atmosphere. 

Realising this connection between convective mixing and the presence
of dust, various models have been developed to describe the cloud
structure (cf.\ the review by Marley et al. \cite{marleyIAU211}). 
The model of Tsuji \cite{tsuUniCloud} assumes a fixed grain
size and defines a lower limiting temperature $T_\mathrm{crit}$,
below which grains would coagulate too fast to stay suspended. 
The vertical extent of the cloud deck is thus defined by the
condensation temperature and  $T_\mathrm{crit}$, the latter being an
empirical parameter that needs to be adjusted by comparison with 
observations.  

A more detailed approach has been developed by Ackerman \&\ Marley 
\cite{ack01}, which uses a simplified description of the
equilibrium between grain settling and mixing to derive modal grain
size and dust density. Above the convection zone, vertical transport
is assumed to occur only due to turbulent mixing with a fixed eddy
diffusion coefficient. Their model also includes a free parameter 
$f_\mathrm{sed}$ to describe the sedimentation efficiency, which again
is adjusted to fit the cloud thickness to observations. 
They find that observed L dwarf colours can be reproduced well using a
relatively small range in $f_\mathrm{sed}$ \cite{marley02}. 

In the model used in the current version of \texttt{PHOENIX}, 
Allard et al.~\cite{settlIAU211,settlDust} describe
a more detailed calculation of the equilibrium between mixing 
and various condensate growth processes.  Vertical
transport within the convection zone is calculated according to mixing
length theory, and overshoot into the radiative layer is
modelled by assuming a velocity law based on
the 3D-hydrodynamical simulations of Ludwig et al.~\cite{ludwig-Mdwarf}. 
The time scales for condensation, coagulation and coalescence, and the
sedimentation speed of grains, are currently calculated according to
the description of Rossow \cite{rossow78} for cloud microphysics in
planetary atmospheres. The removal of elements from the gas phase by 
condensation and sedimentation is self-consistently taken into account
in the solution of the CE system. 
The main differences from \cite{ack01} are thus the
formulation of a depth-dependent mixing velocity driven by convective
overshoot in the radiative zone, and the use of the microphysical time
scale approach of 
Rossow, which adds a level of physical detail, though it 
risks extrapolating results which have only been tested empirically
in the very different conditions of solar system planets. 
Another model based on the work of Rossow has been published by Cooper
et al.~\cite{cooperClouds}, though they  give no details
on their the mixing velocity law, and have not yet included 
the cloud opacity into the atmosphere structure.  

The most advanced and detailed treatment of cloud microphysics to
date has been presented by Woitke \& Helling 
\cite[cf.\ also C.~Helling, this volume]{BerlinDustIII}. 
Their model provides a kinetic description of turbulent mixing and
dust formation and {precipitation}\index{precipitation} processes, 
also based on an
overshooting velocity field calibrated on hydrodynamical models 
\cite{ludwig-Mdwarf}. At this stage only one species, TiO, is
considered, and the computational complexity of these non-local cloud
formation models has so far prohibited incorporation into a
self-consistent model atmosphere code. 

It is worth noting that all these models except \cite{tsuUniCloud} 
predict the formation of dust grains larger than 1\,{\umu m}, and
ranging up to several 100\,{\umu m}~\cite{cooperClouds}, in some layers
of the atmosphere. IR extinction coefficients in these models are thus
no longer independent of the grain size, and measurements of the near-
and mid-IR SED may provide an observational test of the size
distribution. 

In parallel to the settling into deeper layers 
it is also possible that the cloud deck is breaking up into smaller 
features, with the surface coverage decreasing with $T_\mathrm{eff}$. 
The balance of dust rainout and replenishment by
convective upwelling is certainly conductive to the formation of
unstable cloud structures, and evidence of this might be
found in the very rapid change of the SED and the reappearance of FeH
absorption bands in early T dwarfs \cite{adamLT02}. 
But while observations of both periodic and transient variability in L
and some early T dwarfs clearly indicate the existence, and 
rapid evolution, of surface patterns 
\cite{2001A&A...367..218B,2003MNRAS.346..473K}, no
trend for an increase in  amplitude has been found that
would support a significant change of the cloud fraction towards
the L/T boundary \cite{enochLTvar}. 
The models of Ackerman \&\ Marley require a steady decrease in cloud
coverage to explain the observed colour change at the beginning of the T
dwarf sequence \cite{marleyIAU211}, while the settling code of Allard
et al.~\cite{settlIAU211} is able to reproduce these 
broad-band colours with a homogeneous cloud deck, but has not been
tested against all changes of spectral features.  
More detailed comparison with spectroscopic observations will
therefore be required to decide if uniform cloud models can fully
account for the transition from dusty to dust-free atmospheres. 

\subsection{Molecular Abundances and Opacities}\label{molecules}
A serious obstacle to improving models for ultracool atmospheres is
the lack of physical input data for many 
atomic and molecular opacities. Most existing databases of molecular
lines such as HITRAN \cite{hitran98} have
been compiled for planetary atmosphere studies, and are 
oriented towards low temperatures. Consequently, they are highly
incomplete at energy levels strongly populated in brown dwarf and cool
star atmospheres. Stellar atmosphere models therefore rely strongly on
the efforts of laboratory spectroscopists and molecular physicists to
produce comprehensive experimental and theoretical high-temperature
line lists \cite[cf.\ also S.~Johannson, this
volume]{draAtmos,ugjAtmos}.

\begin{figure}[t]
\begin{center}
\includegraphics[width=.9\textwidth]{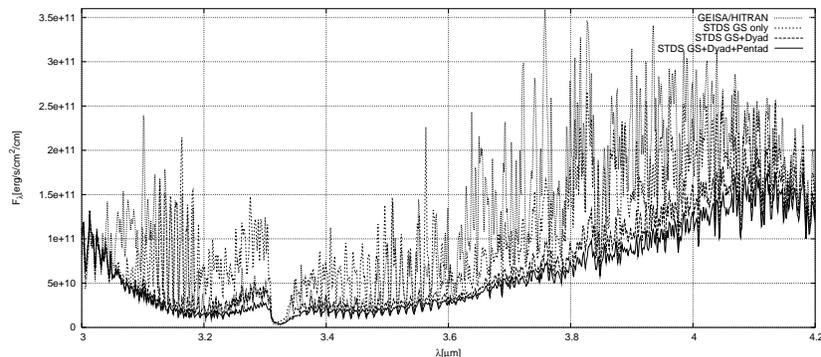}
\end{center}
\caption[]{Theoretical models of the strongest methane absorption
  band for a T dwarf of $T_\mathrm{eff}$\,=\,1000\,K. The plots, as
  labeled from the top, show spectra based on the GEISA/HITRAN line
  list, and on 3 lists calculated with the STDS code~\cite{hahbSTDS}.
} 
\label{lbandSTDS}
\end{figure}

For dense molecular bands  
completeness of all relevant transitions (i.\,e.\ covering all levels
populated at temperatures of 
500\,--\,3\,000\,K), is generally more important to correctly include 
blanketing effects, than accurate individual line strengths and 
positions. 
This is illustrated in Fig.~\ref{lbandSTDS}, which compares lines
from the low-temperature databases with simulations using the STDS
code \cite{rhlcb01,hahbSTDS}, calculated for transitions from
the vibrational ground state, as well as the first (Dyad) and second
(Pentad) excited levels ({\em hot bands}). 
For population numbers typical of brown dwarf conditions, 
lines from each band contribute about the same 
integrated absorption strength, but distributed over approximately
2\,$\times\,10^4$, 2\,$\times\,10^5$, 2\,$\times\,10^6$ and
1.3\,$\times\,10^7$ lines, respectively. As a result, the spectra
calculated from the GEISA/HITRAN lines and from only the ground-state
transitions allow several times more flux to escape between the
lines. The spectra including the hot bands on the other hand 
converge as soon as $\sim\!10^6$ lines are included. 

Completeness and accuracy for the IR water bands in the coolest
brown dwarfs remains problematic \cite{2MASS04pap}. The available data
for ammonia bands, which dominate the mid-IR spectrum in late
T dwarfs, are even more limited. In addition, mounting evidence
suggests that CH$_4$/CO as well as NH$_3$/N$_2$ do not reach their CE
mixing ratios in the cool upper layers of T dwarf atmospheres, as 
some key reaction times could be orders  of
magnitude larger than the timescale for mixing with warmer layers
\cite{saumonIAU211}. 
Probing molecular abundances at low  spectral resolution is
complicated by the strong blending of many bands (e.\,g.\ CO and
H$_2$O at 4.55\,{\umu m}), which might be disentangled by
high-resolution observations. 

\section{Hot Jupiters}\label{planets}

The discovery rate of EGPs has only been slightly behind that of brown
dwarfs. 
Most known EGPs are associated with a parent star  similar to
our Sun and have very small orbital separations compared to
that of Jupiter. Their close proximity to a solar-type star makes the
modeling of such objects  slightly more complicated than brown dwarfs
because both the intrinsic interior source of heat and the intense
extrinsic solar radiation must be accounted for. Barman et
al.~\cite{irradplanets} have modeled EGPs at various orbital distances
and cooling stages. 

\begin{figure}[t]
\begin{center}\hspace*{-1ex}
  \includegraphics[width=.33\textwidth,angle=90]{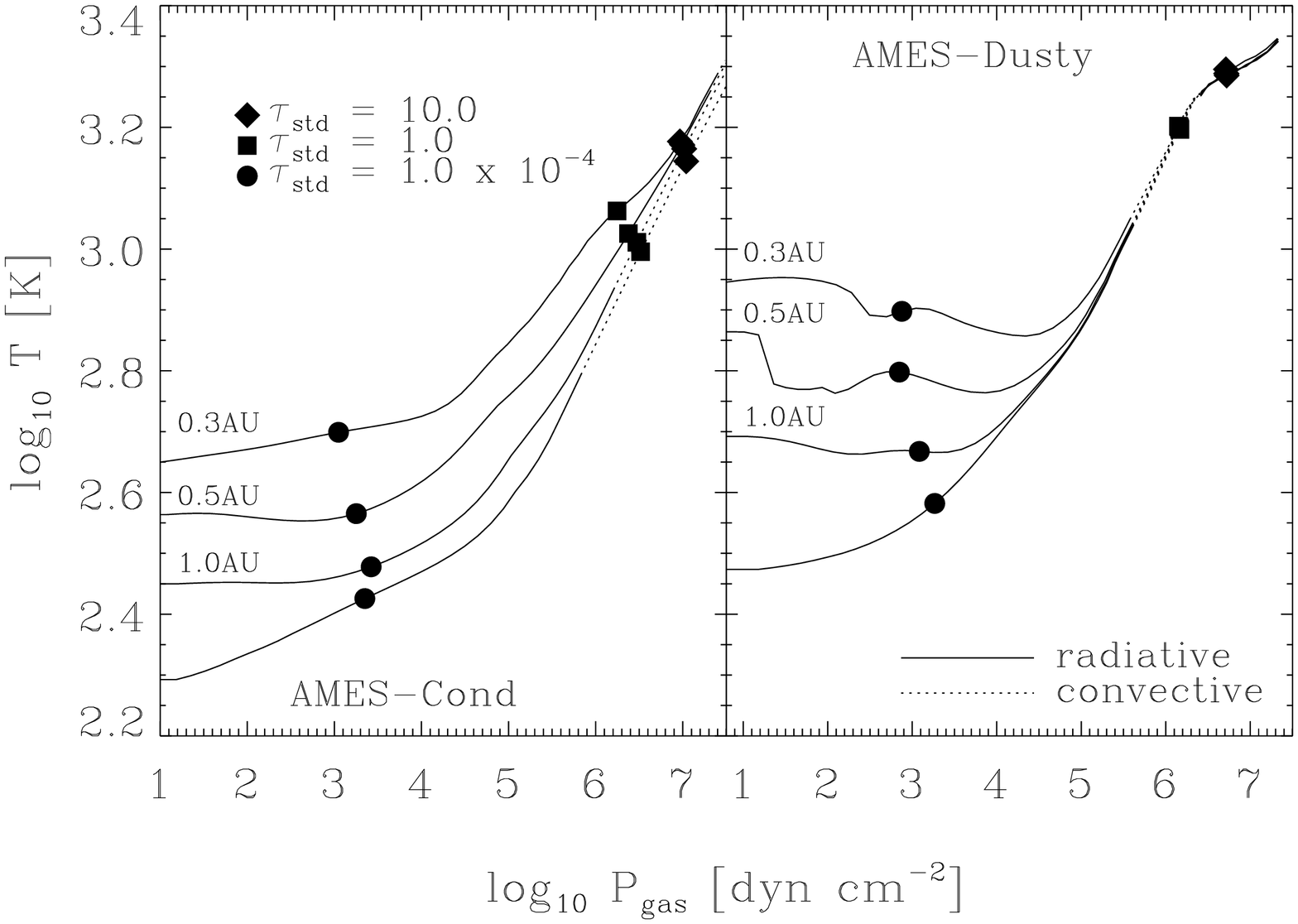}\hspace*{-2.5ex}
  \includegraphics[width=.30\textwidth,angle=90]{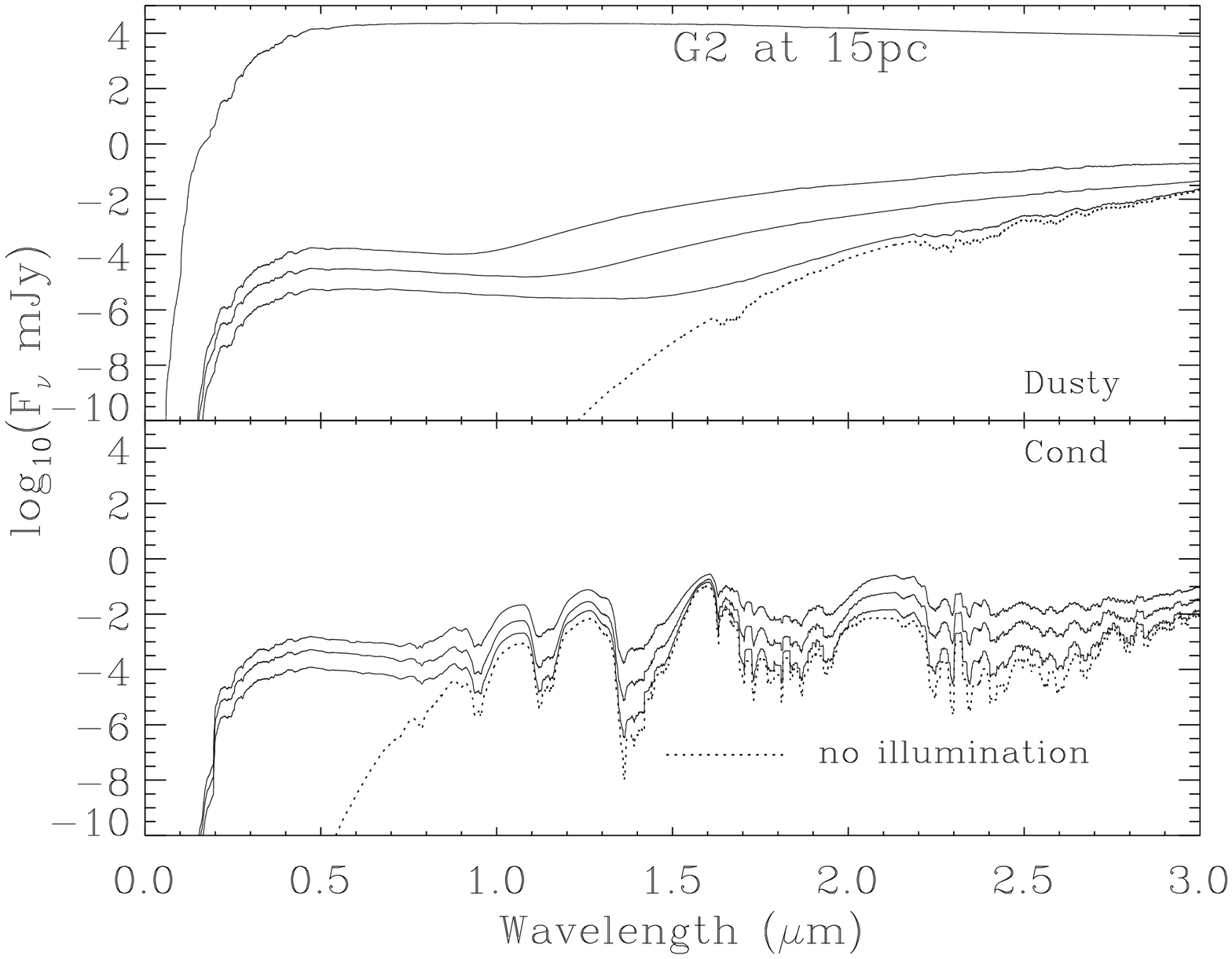}
\end{center}
\caption[]{
  Models for a non-irradiated and irradiated
  ($T_\mathrm{eff}$\,=\,500\,K, log\,$g$\,=\,3.5) planet
  when located 1.0, 0.5 and 0.3 AU from a G2 primary star from
  \cite{irradplanets}. 
  Thermal structures for dust-free and dusty models are shown on the
  left, with the bottom line indicating the case without irradiation. 
  The filled symbols refer to different optical depths ($\tau$)
  defined at $\lambda = 1.2$\,\umu m. 
  The corresponding spectra are plotted on the right. 
  For comparison, the spectrum of the G2V star is
  also shown. All fluxes have been scaled appropriately for the size
  of the planet and primary at a distance of 15 parsecs.}
  \label{irradplanets}
\end{figure}

Figure~\ref{irradplanets} demonstrates the effects of irradiation on
the atmosphere of an EGP which has cooled internally to 500\,K. This
corresponds to an age of about 10 million years
\cite{evolPlanets}. This particular case illustrates the 
dramatic heating that occurs in the atmosphere of EGPs that are found
as close as 0.05 AU from their parent star (hereafter \lq \lq Hot
Jupiters''). Stellar radiation is capable of warming the planet's 
surface to an {equivalent temperature}\index{equivalent temperature} 
above 1600\,K, i.\,e.\ conditions similar to dusty L dwarfs. 
Therefore, depending upon the atmospheric composition
history, we also may expect to find magnesium silicate clouds at the
surface of \lq \lq Hot Jupiters''.  
These models also demonstrate the effects of
{irradiation}\index{irradiation} for
the extreme cases of condensation in an EGP atmosphere. We 
compare the temperature structures for a cloud-free and cloudy
planet located at various orbital separations. In the
cloud-free model, the stellar radiation penetrates deep into the
atmosphere and heats the layers near the radiative-convective
boundary. At small orbital separations, the heating is sufficient to
suppress convection in most of the atmosphere. In the cloudy case, the
dust effectively shields the inner layers of the atmosphere and
heating only occurs at the top of the atmosphere.  
The cloud-free EGP atmospheres have spectra similar to the dust-free, 
free-floating brown dwarf shown as the lowermost curve in
Fig.~\ref{MLTspectra}. However, as the planet is brought closer to the
parent star, the molecular absorption features weaken, due to a
flattening of the temperature profile. The dusty atmospheres have
nearly featureless spectra due to the \lq \lq grey'' characteristics of dust
opacity. When 0.05\,AU is reached (not shown), the upper atmosphere is
thermalised and the near-IR spectrum featureless in both
cases. However, these models  represent the substellar
point (i.\,e.\ the point closest to the star) and therefore only
predict a small portion of the complete observable spectrum. 

\begin{figure}[t]
\begin{center}
\includegraphics[width=.8\textwidth,height=.36\textwidth]{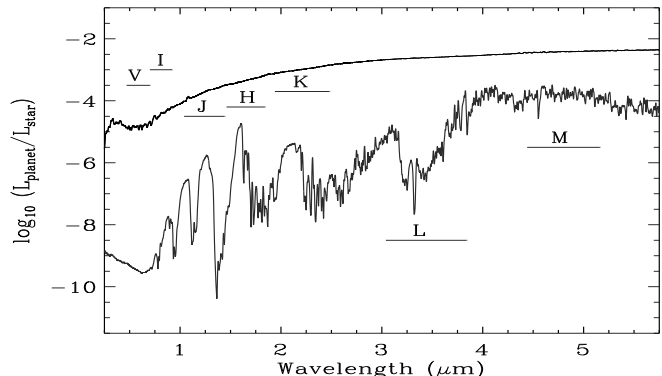}
\end{center}
\caption[]{
  The luminosity ratio for a 1\,$M_\mathrm{J}$ EGP located 0.05\,AU
  (top curve) and 5.0\,AU (bottom curve) from a G2V main sequence
  star. In both cases, the EGP radius is 1.42\,$R_\mathrm{J}$ and
  the intrinsic effective temperature was 500\,K, typical of a
  young planet (a few Myrs) \cite{evolPlanets}. The location of
  various photometric bands are also shown. The contrast is most
  favourable beyond 4\,\umu m.}
  \label{planetsflux}
\end{figure}

It is very likely that a large day/night temperature difference will
exist leading to a horizontal temperature gradient. Thus, 
the observable hemisphere of an EGP may have a complex
spectrum composed of regions with varying degrees of stellar
illumination and a fraction of the night side, depending on the phase
and inclination. The exact thermal profile will also depend on the
global {atmospheric circulation}\index{atmospheric circulation} that would
redistribute the irradiated energy \cite{gui51PegEvol}. 
On the other hand, for orbital separations larger 
than about 1 AU the stellar flux no longer adds significantly to the
thermal near-IR EGP spectrum, implying that such EGPs can be well
approximated by a non-irradiated atmosphere. Such an
approximation is only valid for the thermal radiation emitted beyond 
1\,{\umu m}. At shorter wavelengths, the planet's atmosphere reflects the
stellar flux, making it appear much brighter than isolated
brown dwarfs. The reflected optical starlight persists out to large
orbital distances, as cooler atmospheres form highly reflecting cloud
layers (ices with albedos of 0.5). 

Since planets have been most often found
close to a star, it is important to identify the most favourable
wavelengths for the direct detection of light from those planets. 
Figure~\ref{planetsflux} presents the monochromatic luminosity ratio
of an EGP at 0.05 and 5\,AU from a G2V star. Independent of the
spectral type of the primary (from G to M), the EGP at 0.05\,AU has a
contrast reaching $10^{-2}$ in the $L^\prime$ band. However, when the
planet is at 5\,AU, the contrast reaches only $10^{-3.5}$ at 
5\,{\umu m}, decreasing to $\sim\!10^{-5}$ in the $H$ bandpass. 
Note that these numbers are, for reasons above, optimistic 
estimates and assume a \lq \lq full moon'' planetary phase. The contrast turns
up at optical wavelengths due to starlight reflected by the
planet. The night side of a \lq \lq Hot Jupiter'' may look more like the EGP
with an orbital separation of 5 AU, except of course that there would
be no reflection effects, with such a large day/night difference
likely  producing strong temporal variations of
the contrast as the planet orbits the star. 

\section{Conclusions}
Ten years of observational and theoretical work have brought us 
thorough understanding of the basic physical 
properties of brown dwarfs. While many details, especially regarding
non-equilibrium processes, and temporal and spatial variability of the
atmosphere, still need to be addressed, current models provide a solid
foundation for understanding extrasolar giant planets, and predicting
their observable signatures. 
BD and EGP 
(unirradiated as well as \lq \lq Hot Jupiter''-like) model atmospheres, 
thermal profiles, synthetic spectra, photometry, and 
evolution calculations are available for all evolutionary stages
from a few Myrs from our web site at:
\href{http://perso.ens-lyon.fr/france.allard/}{http://perso.ens-lyon.fr/france.allard/}.

\subsection*{Acknowledgments}
{\small
D.~H.\ wishes to thank Ulli K{\"a}ufl and ESO for the invitation to
this workshop, and Christiane Helling and Bengt Gustafsson for
instructive  comments on the presentation. This research was supported
by the United States National Science Foundation under grant N-Stars
RR185-258, and by NASA under grant NLTE RR185-236. 

We are much indebted to the US DoE/NERSC and the German HLRN for
generous allocation of CPU time at their supercomputing facilities. 
}

\end{document}